\documentstyle{mn}
\newcommand{\mc}{\multicolumn}
\newcommand{\ph}{\phantom}

\newcommand{\ti}{\times}

\newcommand{\aip}{{\small ${\cal AIPS}$}}
\newcommand{\gtsim}{\mbox{{\raisebox{-0.4ex}{$\stackrel{>}{{\scriptstyle\sim}}
$}}}}

\newcommand{\figstart}[1]
    { \begin{figure}[htb]
      \begin{picture}(0,#1) }
\newcommand{\figend}[4]
    { \end{picture}
      \special{#1}
      \caption[#2]{#3}
      \label{#4}
      \end{figure} }

\begin{document}

\title[The central engines of radio-quiet quasars]{The central engines
of radio-quiet quasars}

\author[Blundell \& Beasley]{Katherine M.\,Blundell$^1$ and Anthony J.\ 
Beasley$^{2}$\\ 
$^1$University of Oxford Astrophysics, Keble Road, Oxford, OX1 3RH, U.K. \\
$^2$National Radio Astronomy Observatory, Socorro, NM 87801, U.S.A. \\ }

\maketitle

\begin{abstract}

Two rival hypotheses have been proposed for the origin of the compact
radio flux observed in radio-quiet quasars (RQQs). It has been
suggested that the radio emission in these objects, typically some two
or three orders of magnitude less powerful than in radio-loud quasars
(RLQs), represents either emission from a circumnuclear starburst or
is produced by radio jets with bulk kinetic powers $\sim 10^3$ times
lower than those of RLQs with similar luminosity ratios in other
wavebands.  We describe the results of high resolution ($\sim $
parsec-scale) radio-imaging observations of a sample of 12 RQQs using
the Very Long Baseline Array (VLBA). We find strong evidence for
jet-producing central engines in 8 members of our sample.

\end{abstract}


\section{Introduction \& Background}

The bimodality of radio luminosity in the quasar population (Miller,
Peacock \& Mead 1990) poses many questions about the fundamental
nature of the emission mechanisms in Radio-Quiet Quasars (RQQs) and
Radio-Loud Quasars (RLQs). The total radio luminosity is typically two
or three orders of magnitude lower for a RQQ than for a RLQ (Miller,
Rawlings \& Saunders 1993) with similar luminosity ratios in all other
wavebands.  The double, often co-linear, morphology exhibited by many
RLQs on arcsecond scales (e.g., Bridle et al.\ 1994) usually comprises
a bright central core with hotspots at the outermost edges of the
radio structure. In addition, lobes of {\em extended emission} are
seen which are fed by jets, via backflow out of the hotspots. In
contrast, radio images of RQQs often show merely a weak component,
coincident with the optical quasar nucleus, which in some cases is
resolved (Miller et al.\ 1993).

It has been proposed that the activity in RQQs is supplied by a
`starburst', i.e., strongly radiative supernovae and supernova
remnants (SNRs) in a very dense environment (Terlevich et al.\
1992). Sopp \& Alexander (1991) argued this on the basis of the
striking continuity between the far-infra red --- radio correlation of
RQQs and that of star-forming galaxies, ultra-luminous infra-red
galaxies and also Seyfert galaxies. This is offset from the same
correlation for RLQs and radio galaxies. Alternatively, if the energy
supply arises from accretion onto a massive black hole, the radio
emission from RQQs (as in RLQs) is caused by radio jets, but the bulk
kinetic powers of these jets are for some reason $\sim 10^3$ times
lower than those of RLQs (Miller et al.\ 1993).

To address the question of whether the radio emission in radio-quiet
quasars is associated with starbursts or with weak-jet producing
central engines, we have undertaken a programme of imaging a sample of
RQQs using VLBI techniques. This provides a definitive test between
the two rival hypotheses for the radio emission in a given RQQ, since
one can derive a size scale (from milli-arcsecond resolution
measurements) from which the observed luminosity is emitted and
compare this with the size scales on which SNRs are found to be
distributed. Moreover, a mere detection with VLBI implies the
brightness temperature of the emission $T_{\rm B}
\stackrel{>}{_{\sim}} 10^6$K, while typical supernovae/supernova
remnants have $T_{\rm B} \stackrel{<}{_{\sim}} 10^5$ K (Muxlow et al.\
1994).

We previously reported an experiment to image a well-known nearby
radio-quiet quasar (E\,1821+643) with the VLBA (Blundell et al.\
1996). In detecting it at 5 and 8 GHz on milli-arcsecond scales, we
found a high brightness temperature ($T_{\rm B} \sim 10^{9}\ {\rm K}$)
which precluded the possibility of star-formation as the origin of its
radio emission.  It was instead consistent with a mechanism similar to
the central engines postulated for radio-loud quasars. In order to
establish whether this result was typical of the radio-quiet quasar
population we performed detection experiments using the VLBA on a
sample of 12 RQQs, the results of which we present here.

We describe in Section 2 of this paper the details of our sample
selection and include a discussion of the conventional criteria used
to classify whether a particular quasar is radio-loud or
radio-quiet. In Section 3 we describe our observing method and we
summarise our results. In Section 4 we discuss the physical
implications of our results for the models discussed earlier; we also
explore when a quasar is appropriately classified as radio-quiet and
consider what may be the counterparts of the RQQ population which
undergo Doppler boosting.


\section{Sample selection}

Our goal was to select a sample of targets which were capable of being
imaged with the VLBA and representative of the RQQ population. We used
the Bright Quasar Survey (BQS) (Schmidt \& Green 1983) which has been
studied on arcsecond scales with the Very Large Array (VLA) and also
optically (Miller et al.\ 1993, Kellermann et al.\ 1994).  It is
necessary to impose a flux-density limit to allow a reasonable chance
of detection: if the core flux density of a RQQ is known to be $\ll 2$
mJy, then the source will not be detected in VLBA observations, even
if it is extremely compact, and we would learn nothing.  From the
sample studied by Miller et al.\ (1993) we selected those quasars
whose core had a peak flux density at 5 GHz (measured using the VLA in
A-array) $ \geq 2$ mJy/beam. We then selected any other quasars from
the sample of Kellermann et al.\ (1994) which were not in the sample
of Miller et al.\ and had a total flux density measured in VLA D-array
to be $\geq 2$ mJy.  It was then necessary to eliminate RLQs by
applying criteria which identify a quasar as radio-loud. There exist a
number of (largely overlapping) criteria which delimit the RLQ
population from the RQQ population: (1) a ratio $R$, of radio to
optical emission \gtsim\ 10 has been taken to indicate that the quasar
is radio-loud (Kellermann et al.\ 1994). (2) If the radio
luminosity\footnote{The cosmology which has been assumed in this paper
is that ${\em H}_{\circ} = 50\,{\rm km\,s^{-1}\,Mpc^{-1}}$, ${\em
q_{\circ}} = 0$ and $\Lambda = 0$.}  is above $10^{25}\,{\rm
W\,Hz^{-1} }$ (Kellermann et al.\ 1994) the quasar is deemed to be
radio-loud. (3) The distinction between radio-loud and radio-quiet
quasars is apparent in the narrow-line luminosity versus radio
luminosity plane (Miller et al.\ 1993) where a clear gap in radio
luminosity separates the $z < 0.5$ RQQs from the $z < 0.5$ RLQs. Of
the radio-quiet quasars, Miller et al.\ (1993) identified a sub-set
they termed `radio-intermediate quasars' (RIQs) with luminosities at 5
GHz between $\sim 10^{23}$ and $10^{24}\, {\rm W\,Hz^{-1}\,sr^{-1}}$
(see Section 4.3); note the different units used by these authors
compared to Kellermann et al.\ (1994).

A quasar was retained in our sample if it was radio-quiet according to
at least one of the three above classifications.
Table~\ref{tab:source} lists the twelve objects selected, and a key in
this table indicates which of the three criteria described above led
to the inclusion of a particular quasar.  Two other objects met these
conditions but were excluded from the observing programme for the
following reasons: for 1138+04, the offset between the optical quasar
position and radio emission of 24 arcsec (Kellermann et al.\ 1994)
implied that this radio emission was not that of the core; 1634+70 was
excluded due to scheduling constraints.

\begin{table*}
\begin{tabular}{lllcrcllllll}
\hline	      									                                      
\mc{1}{c}{PG name} &\mc{1}{c}{Alternative} & \mc{1}{c}{Key}  & \mc{3}{c}{VLA 5 GHz}   &\mc{3}{c}{Redshift} &\mc{1}{c}{On-source} & \mc{1}{c}{Phase} & \mc{1}{c}{Date}        \\
                   &\mc{1}{c}{name}        &                 & \mc{3}{c}{flux density}&\mc{3}{c}{ }        &\mc{1}{c}{time (mins)}& \mc{1}{c}{calibrator} &            \\[0.2cm]
0003+199           & Mkn 335               & PRQ             && 3.1$\ph{^{\dag}}$      &&   & 0.025 &  &  48                 & 0007+171   &96 Jun 09   \\
0007+106           & III Zw 2              & \ph{PRQ}I       && 155.0$\ph{^{\dag}}$    &&   & 0.089 &  &  88                 & self       &96 Jun 09   \\
0157+001           & Mkn 1014              & PRQ             && 6.0$\ph{^{\dag}}$      &&   & 0.164 &  &  48                 & 0215+015   &96 Jun 09   \\
0923+129           & Mkn 705               & PRQ             && 2.8$\ph{^{\dag}}$      &&   & 0.029 &  &  92 + VLA           & J0921+1350 &96 Sep 22   \\
1116+215           & \mc{1}{c}{--}         & PRQ             && 2.0$\ph{^{\dag}}$      &&   & 0.177 &  &  69 + VLA           & J1125+2005 &96 Sep 22   \\
1216+069           & \mc{1}{c}{--}         & PR\ph{Q}I       && 5.0$\ph{^{\dag}}$      &&   & 0.334 &  &  84                 & 1219+044   &96 Jun 15   \\
1222+22            & \mc{1}{c}{--}         & \ph{P}R\ph{QI}  && 11.8$^{\dag}$          &&   & 2.046 &  &  48                 & 1222+216   &96 Jun 09   \\ 
1309+355           & \mc{1}{c}{--}         & P\ph{RQ}I       && 54.1$\ph{^{\dag}}$     &&   & 0.184 &  &  48                 & 1315+346   &96 Jun 09   \\
1351+640           & \mc{1}{c}{--}         & PRQ             && 20.0$\ph{^{\dag}}$     &&   & 0.088 &  &  48                 & 1342+662   &96 Jun 09   \\
1407+26            & \mc{1}{c}{--}         & \ph{P}R\ph{QI}  && 7.9$^{\dag}$           &&   & 0.94  &  &  84                 & 1404+286   &96 Jun 15  \\ 
1700+518           & \mc{1}{c}{--}         & PR\ph{Q}I       && 2.2$\ph{^{\dag}}$      &&   & 0.292 &  &  69 + VLA           & J1705+5109 &96 Sep 22   \\
2209+184           & II Zw 171             & P\ph{RQ}I       && 117.0$\ph{^{\dag}}$    &&   & 0.07  &  &  48                 & 2209+236   &96 Jun 09   \\ 
\hline
\end{tabular}
\parbox{162mm} {\caption[sourcetable]{\label{tab:source} Observed VLBA
RQQ sample, including alternative names (if in common usage) and a key
classifying the sources as follows: P indicates that the 5-GHz
luminosity $< 10^{25}\ {\rm W\,Hz^{-1}}$; R indicates that $R < 10$; I
indicates RIQ while Q = RQQ, as designated by Miller et al.\ (1993).
The fifth column lists the core flux density at 5 GHz measured with
the VLA in A-array by Miller et al.\, except for those sources marked
$^{\dag}$ where this value is the integrated 5-GHz flux density
(measured by Kellermann et al.\ 1994). Redshifts of 0157+001 to
1351+640 (except for 1222+22) as listed above were taken from Miller
et al.\ (1992). The redshift of 1407+26 is quoted from McDowell et
al.\ (1995); the redshifts of 0003+199, 1222+22, 1700+518 and 2209+184
are quoted from Schmidt \& Green (1983); the redshift of 0007+106 is
quoted from Sargent (1970). This table also lists the on-source time
in minutes, indicates those sources for which the VLA was used in
phased-array mode, the phase calibrator used for the phase
referencing, and the date of the quoted observations.  }}
\end{table*}

\section{Observations \& Results}

All targets were observed with the 10-antenna VLBA of the National
Radio Astronomy Observatory at 8.42 GHz. With the exception of
0007+106 all targets were observed in a phase-referenced mode (Beasley
\& Conway 1995), i.e. frequent observations of an adjacent bright
source were made to provide phase corrections for the interferometer
array. For those targets with low flux density (as indicated in
Table~\ref{tab:source}) we used the VLA in phased array mode for extra
sensitivity.  The data were processed using the VLBA correlator which
generated four 8\,MHz continuum channels in the four Stokes
parameters.  Synthesis imaging of the data was performed using the
NRAO \aip\ system.

\begin{figure*} 
\centering
\begin{picture}(162,595)(0,0)
\put(-190,330){\includegraphics{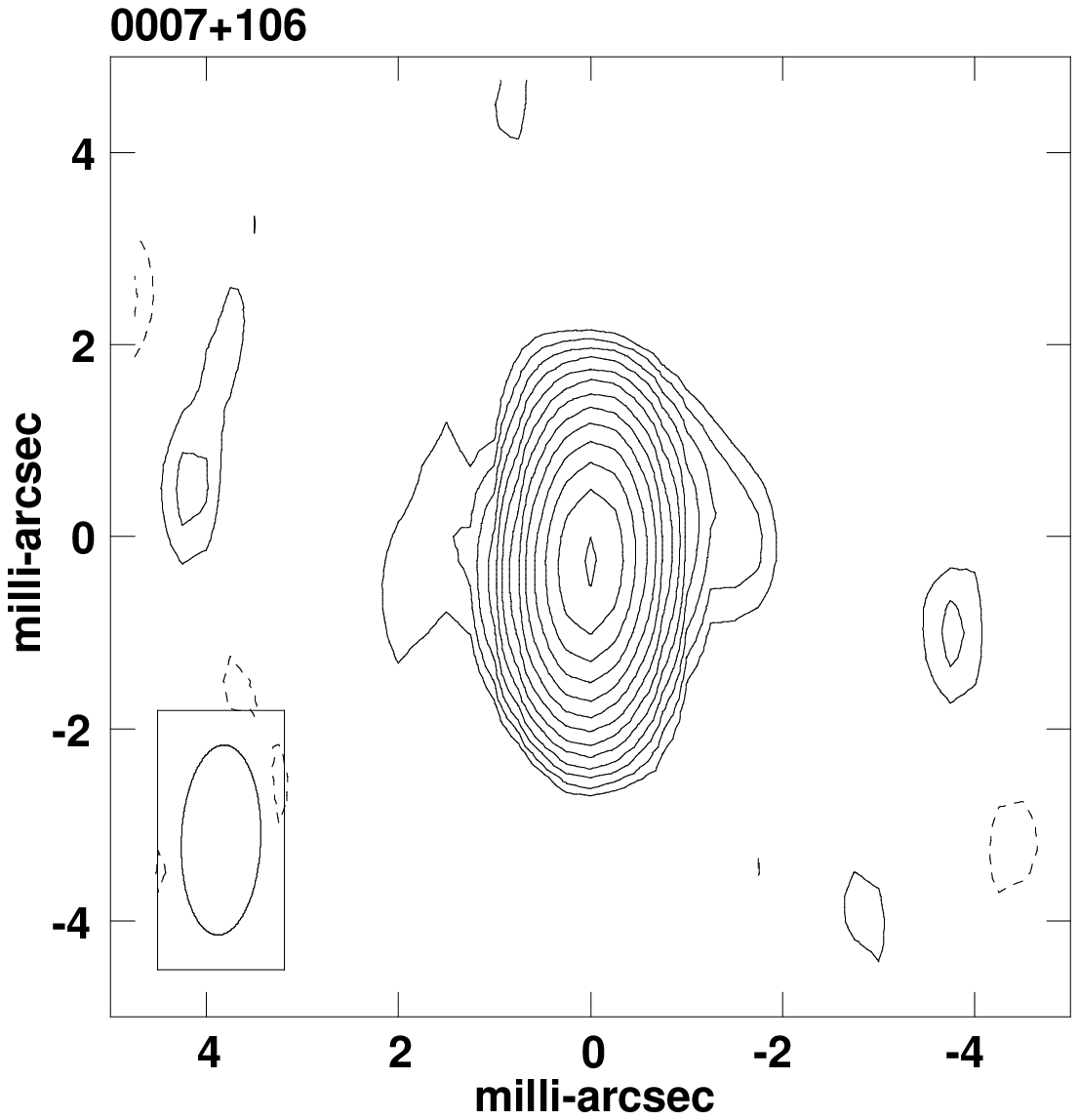}}
\put(21,330){\includegraphics{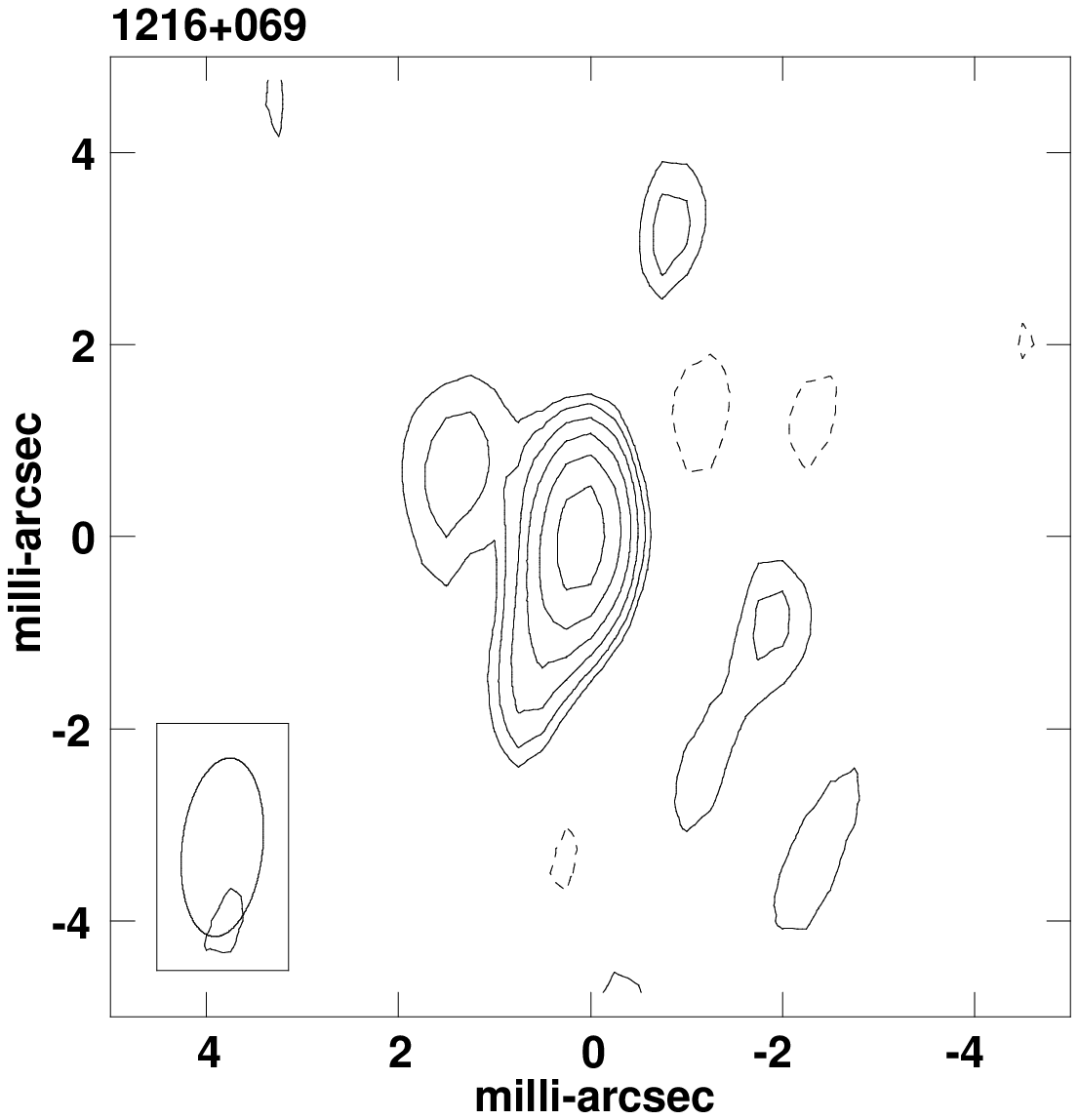}}
\put(-190,180){\includegraphics{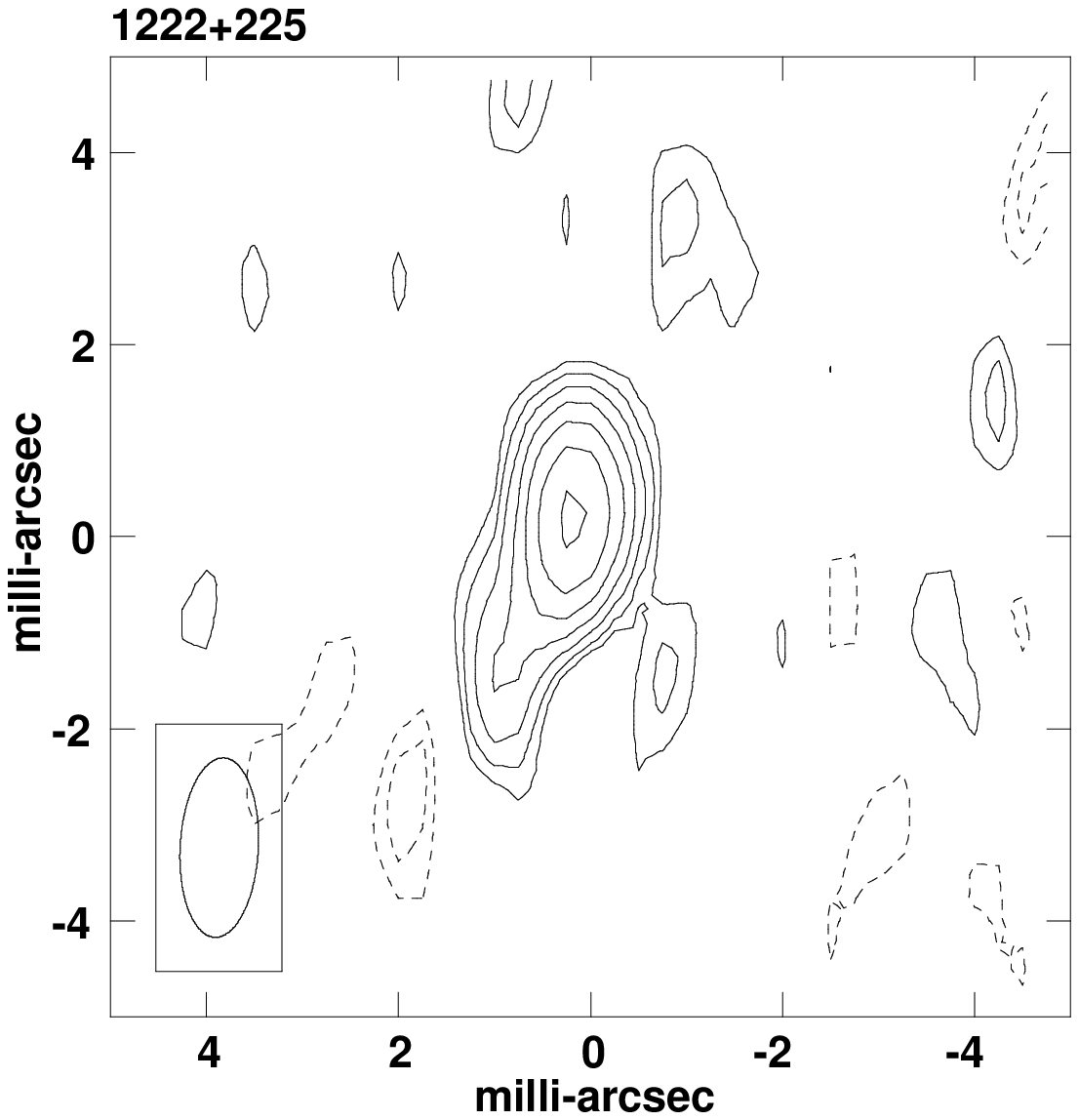}}
\put(21,180){\includegraphics{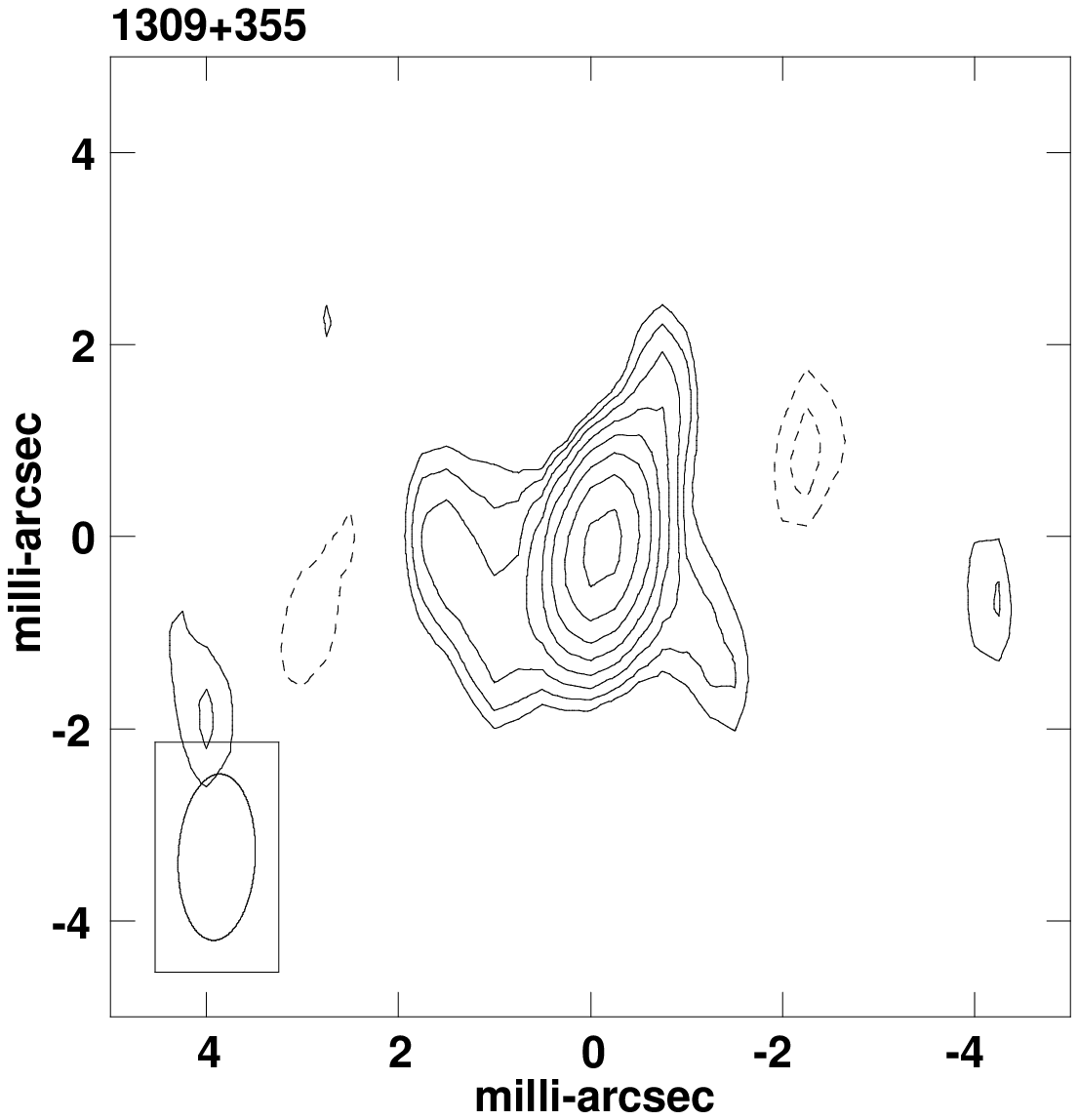}}
\put(-190,30){\includegraphics{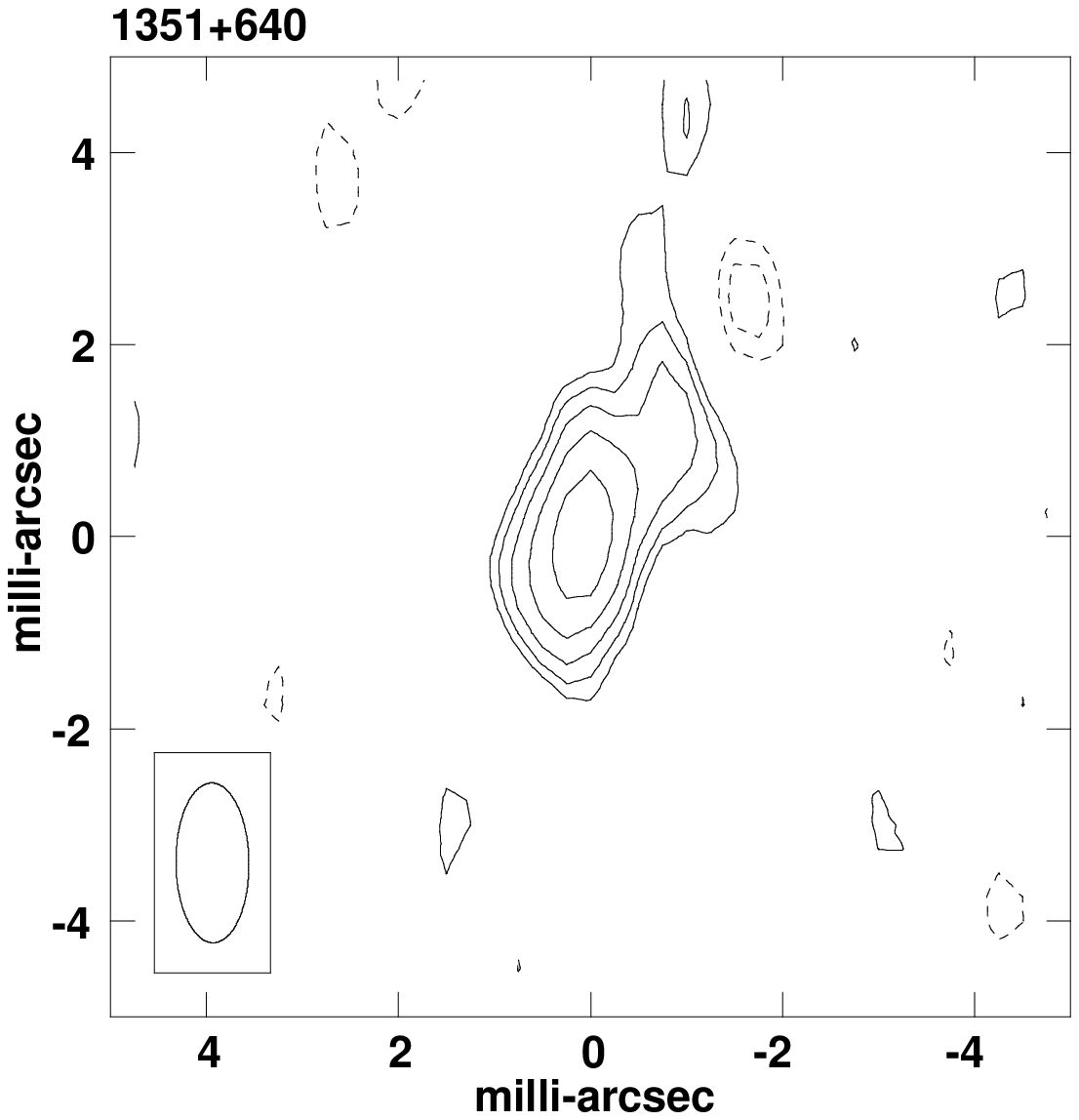}}
\put(21,30){\includegraphics{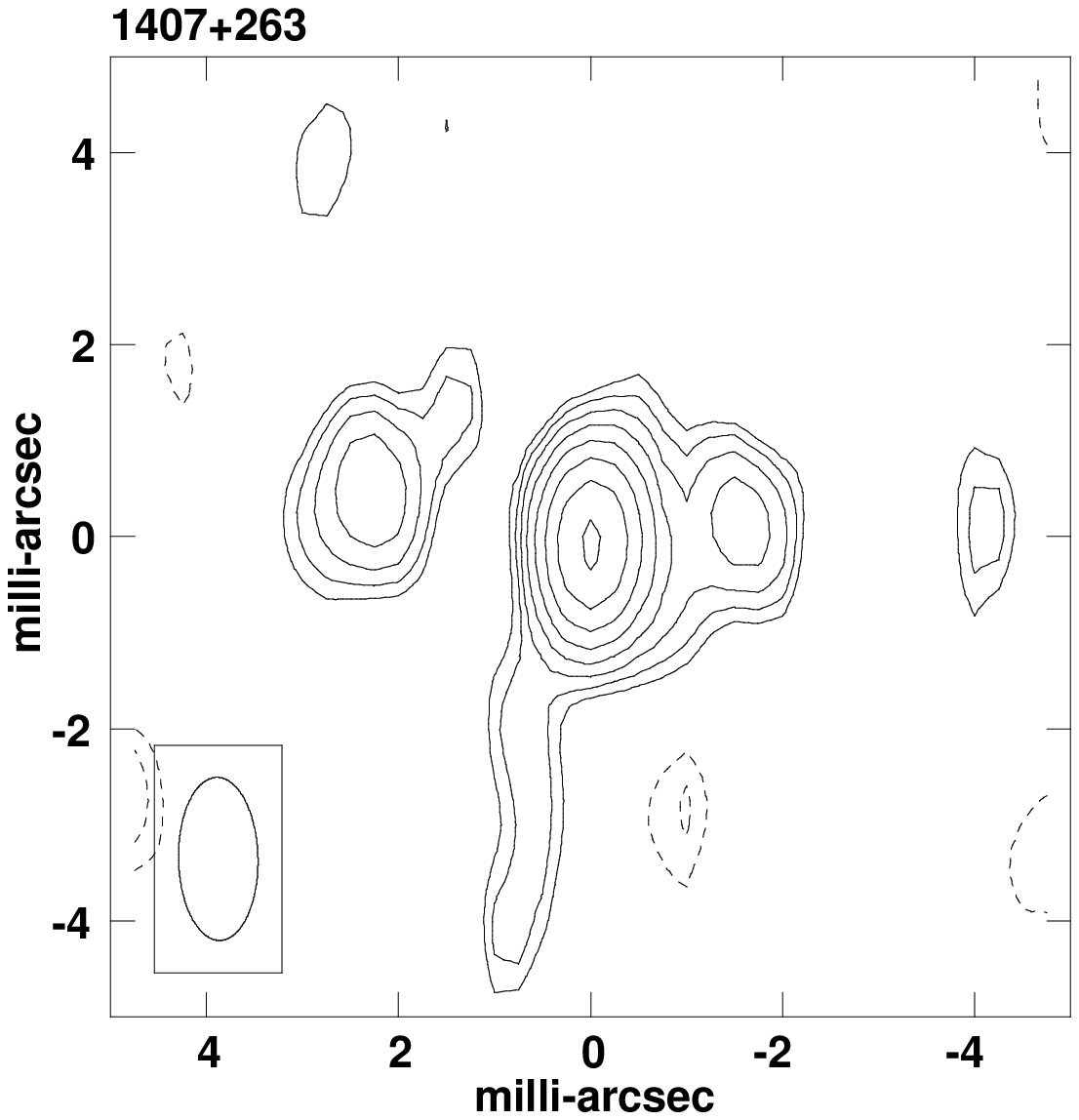}}
\put(-190,-120){\includegraphics{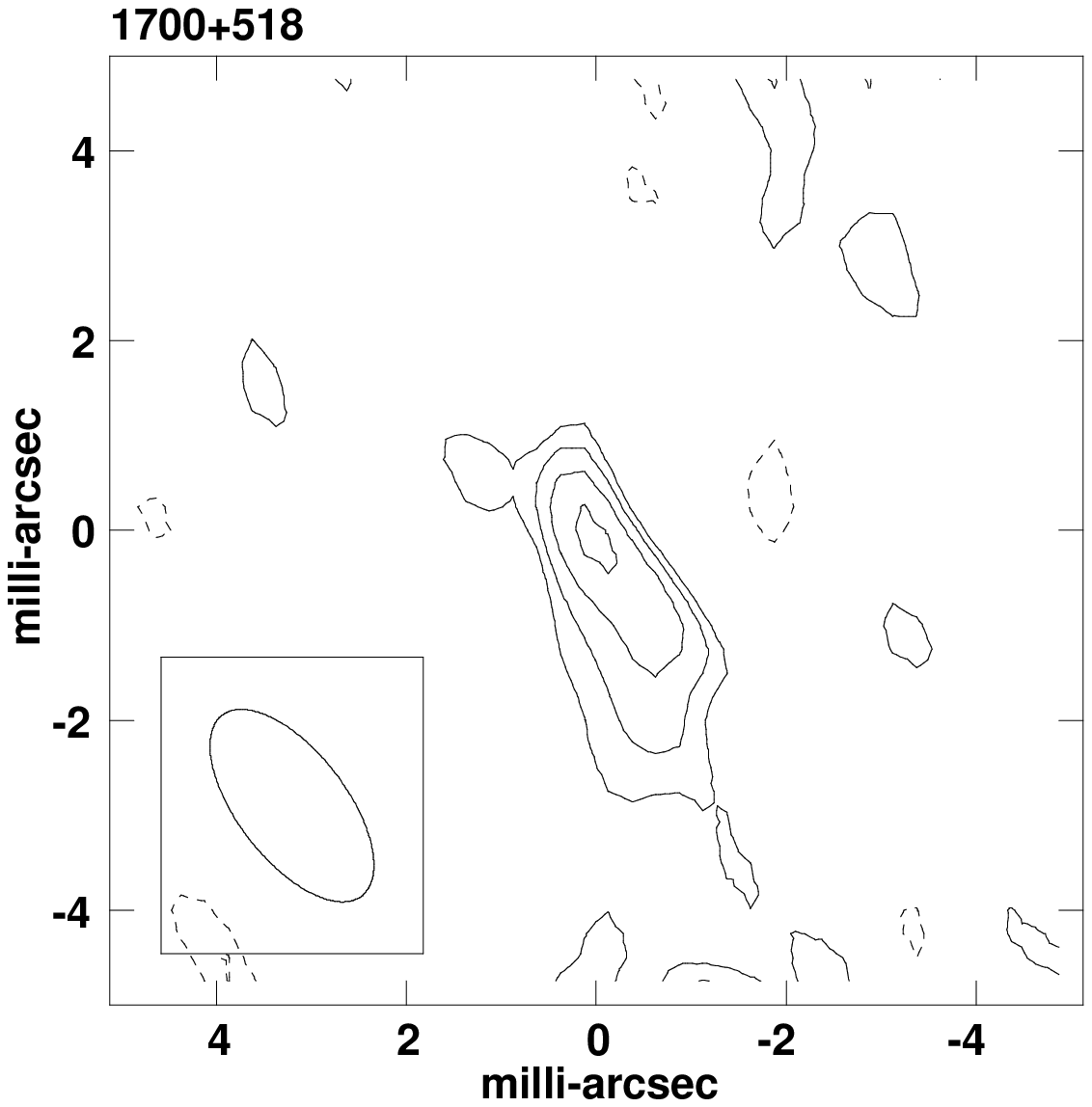}}
\put(21,-120){\includegraphics{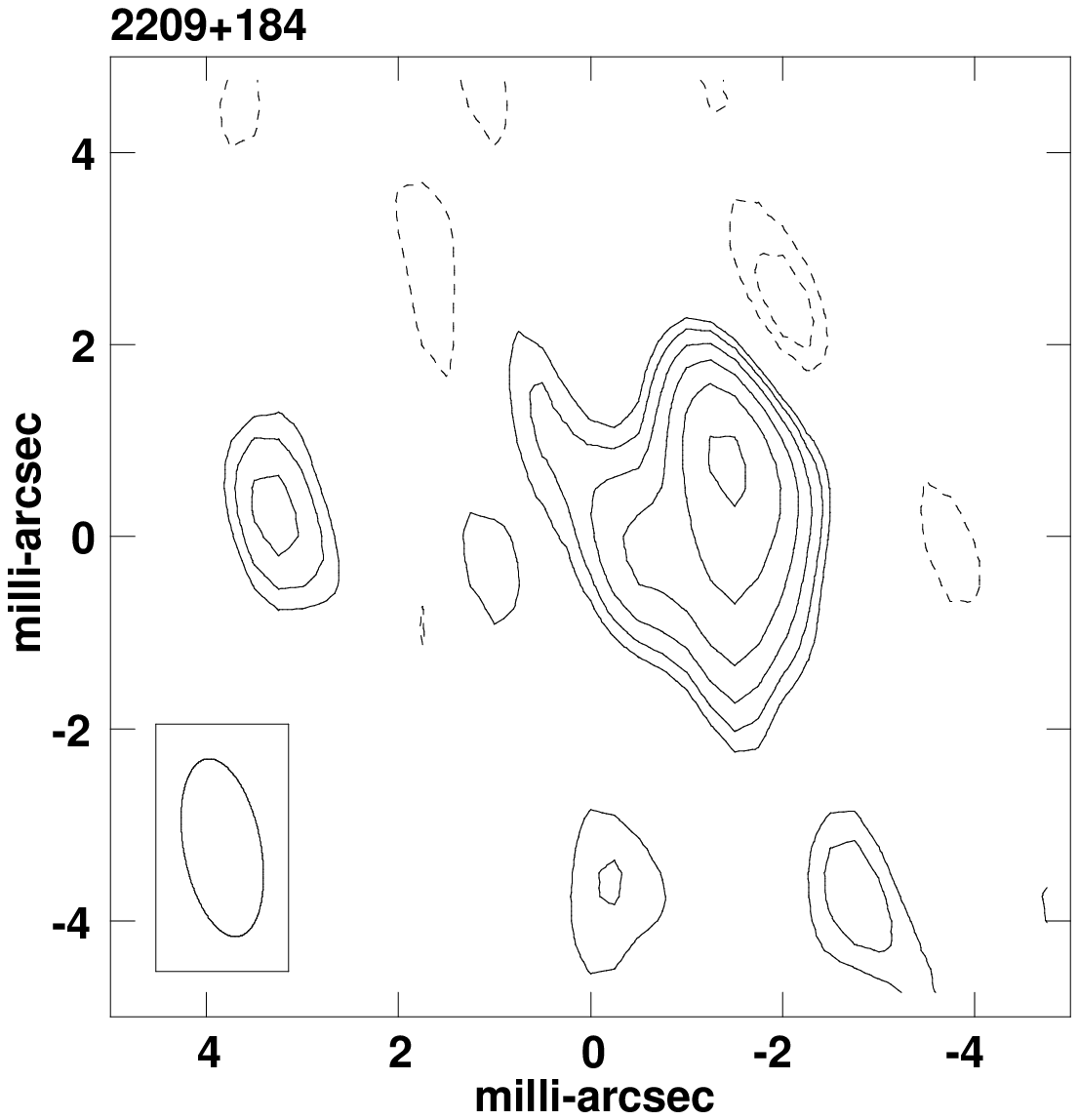}}
\end{picture}
\parbox{162mm}{ \caption[junk]{\label{fig:maps} VLBA images at 8.4
   GHz. The lowest solid contour levels in each case are: 0007+106,
   0.87 mJy/beam; 1216+069, 0.24 mJy/beam; 1222+22, 0.25 mJy/beam;
   1309+355, 1.2 mJy/beam; 1351+640, 0.28 mJy/beam; 1407+26, 0.20
   mJy/beam; 1700+518, 0.15 mJy/beam; 2209+184, 12.0 mJy/beam. Other
   solid contours are at $\sqrt{2}$, 2, 2$\sqrt{2}$\ldots times these
   values, while dashed contours are plotted at $-\sqrt{2}$ and $-1$
   times these values.  Extended structures indicative of jets are
   seen some of the maps although at these low brightness levels, and
   with the short snapshot observations used in this detection
   experiment, it is difficult to comment on the reality of these
   features.  }}
\end{figure*}

The results of our observations are presented in
Table~\ref{tab:meas_results}. We detected 8 out of 12 of the quasars
in our sample with brightness temperatures between $10^7$ and $10^9$ K
derived according to the following formula: $T_{\rm B} = \lambda^2
S_{\lambda} (1 + z) / (2 k_{\rm B} \Omega )$, where $T_{\rm B}$ is the
brightness temperature in Kelvin, $\lambda$ is the wavelength in
metres, $S_{\lambda}$ is the flux density in Jansky at wavelength
$\lambda$, $k_{\rm B}$ is the Boltzmann constant, $\Omega$ is the
solid angle subtended by the emitting region and $z$ is the redshift.
Images of all sources we detected are shown in Figure~\ref{fig:maps}.
We see structures indicative of jets in sources 1216+069, 1222+225,
1351+640 and 1407+26 although at these low brightness levels, and with
the short snapshot observations used in this detection experiment, it
is difficult to ascertain the reality of these features; more detailed
observations of one of these sources are underway. The detection of
1700+518 is only at $3\sigma$ significance, and we consider this
detection marginal.  We present the luminosity and a characteristic
physical size of the emitting region for each detected source in
Table~\ref{tab:der_results}, according to the cosmology assumed
throughout this paper.

\subsection{Comparison of VLA and VLBA flux densities}

Although we are not in possession of multi-epoch radio images on these
milli-arcsec scales, comparison of our integrated VLBA flux densities
at 8.4 GHz with VLA flux densities at 5 GHz measured over a decade ago
(Miller et al.\ 1993, Kellermann et al.\ 1994) indicate differences of
only a factor of a few, which can be interpreted as arising from the
resolving out of some extended emission, in combination with spectral
differences and intrinsic radio variability.  In no case does the 8.4
GHz flux density measured with the VLBA in 1996 exceed the
measurements made at 5 GHz in the 1980s by Miller et al.\ and
Kellermann et al.  We are able to make a comparison between the
simultaneous VLA and VLBA measurements of flux densities for those 3
sources which were observed with the VLA in phased array, and
indicated in Table~\ref{tab:meas_results}.  For the two objects which
we did not detect with the VLBA but which we did observe with the VLA
the flux densities are very low (4 mJy for 0923+129 and 1.5 mJy for
1116+215).  For now we proceed with the question of the properties of
the central engine which must be present in RQQs to sustain, over many
years, the luminosities quoted in Table~\ref{tab:der_results}.

\section{Discussion}
\subsection{Implications}

Our high detection rate from these snapshot observations supports the
findings of our initial experiment on the RQQ, E1821+643 (Blundell et
al.\ 1996). The 4 RQQs which we did not detect in these observations
are those with the lower VLA flux densities (see
Table~\ref{tab:source}).

\begin{table*}
\begin{tabular}{lllrccrcl}
\mc{1}{c}{PG name} & \mc{1}{c}{R.A.}   & \mc{1}{c}{Dec.}   & \mc{2}{c}{VLBA}     & \mc{1}{c}{Synthesised}     & \mc{1}{c}{Total flux}   & \mc{2}{c}{Total flux}   \\
                   & \mc{1}{c}{J2000.0}& \mc{1}{c}{J2000.0}&\mc{2}{c}{8.4 GHz peak}& \mc{1}{c}{beam}          & \mc{1}{c}{density from} & \mc{2}{c}{density from} \\
                   &                   &                   &\mc{2}{c}{(mJy/beam)}& \mc{1}{c}{({\sc fwhm}/mas)}& \mc{1}{c}{VLBA (mJy)}   & \mc{2}{c}{VLA (mJy)}          \\
\hline                                   				       
\hline                                   				       
0003+199 &  \mc{1}{c}{--} & \mc{1}{c}{--}  & $<0.8$  &        &                 &                       &\ph{xxx} &    \\
0007+106 & 00 10 31.00587 & 10 58 29.5037  & 57.7    &        & $1.98 \ti 0.83$ &  $68.3 \pm 1.4$ & &            \\
0157+001 &  \mc{1}{c}{--} & \mc{1}{c}{--}  & $<1.5$  &        &                 &                 & &            \\
0923+129 &  \mc{1}{c}{--} & \mc{1}{c}{--}  & $<0.4$  &        &                 &                 & & 4.1        \\
1116+215 &  \mc{1}{c}{--} & \mc{1}{c}{--}  & $<0.3$  &        &                 &                 & & 1.5        \\
1216+069 & 12 19 20.93171 & 06 38 38.4679  &  1.7    &        & $1.87 \ti 0.84$ &   $2.0 \pm 0.3$ & &            \\
1222+22  & 12 25 27.40090 & 22 35 13.0522  &  2.1    &        & $1.88 \ti 0.80$ &   $3.2 \pm 0.4$ & &            \\
1309+355 & 13 12 17.75278 & 35 15 21.0857  & 14.9    &        & $1.73 \ti 0.80$ &  $23.8 \pm 2.1$ & &            \\
1351+640 & 13 53 15.83069 & 63 45 45.6856  &  1.5    &        & $1.66 \ti 0.76$ &   $2.9 \pm 0.5$ & &            \\
1407+26  & 14 09 23.90866 & 26 18 21.0557  &  2.3    &        & $2.33 \ti 0.98$ &   $4.7 \pm 0.4$ & &            \\
1700+518 & 17 01 24.82093 & 51 49 20.4995  &  0.8    &        & $1.98 \ti 0.98$ &   $0.8 \pm 0.2$ & & 4.9        \\
2209+184 & 22 11 53.88876 & 18 41 49.8634  & 69.8    &        & $1.88 \ti 0.80$ & $179.7 \pm 17.1$& &            \\
\hline\hline								                       
\end{tabular}								                       
\parbox{162mm}{\caption[sourcetable]{\label{tab:meas_results} Results
from the VLBA observations are tabulated: columns 2 and 3 list the
(J2000) Right Ascension and Declination respectively and column 4
lists the peak flux densities which were obtained using the
Gaussian-fitting task {\sc imfit} within \aip\ in units of
mJy/beam. Non-detections are quoted as $< 3\sigma$. Column 6 shows the
total flux density measured by the VLBA in mJy. Column 7 lists the VLA
flux densities measured simultaneously with the VLBA observations for
those sources where the VLA was used in phased array mode.}}
\end{table*}
\normalsize

\begin{table*}
\begin{tabular}{llllccrc}
\mc{1}{c}{PG name} & \mc{1}{c}{$T_{\rm B}$(K)}& \mc{1}{c}{Deconvolved} & \mc{1}{c}{Limit}      &\mc{1}{c}{Log$_{10}(L_{8.4})$}         & \mc{3}{c}{Emitting}    \\ 
                   &                          & \mc{1}{c}{size (mas)}  & \mc{1}{c}{$T_{\rm B}$}&\mc{1}{c}{(${\rm W Hz^{-1} sr^{-1}}$)} & \mc{3}{c}{region (pc)} \\ 
\hline                                                                                                
\hline                                                                                                
0003+199           &                          &                        &                       &     \mc{1}{c}{--}                     & & \mc{1}{c}{--} &          \\ 
0007+106           &  $7.5\ti 10^8$           & $0.4 \ti 0.4$          & $7.7\ti 10^9$         &   23.22                               & & 2.9  &                   \\ 
0157+001           &                          &                        &                       &     \mc{1}{c}{--}                     & & \mc{1}{c}{--}  &         \\ 
0923+129           &                          &                        &                       &     \mc{1}{c}{--}                     & & \mc{1}{c}{--}  &         \\ 
1116+215           &                          &                        &                       &     \mc{1}{c}{--}                     & & \mc{1}{c}{--}  &         \\ 
1216+069           &  $2.8\ti 10^7$           & $0.7 \ti 0.4$          & $1.6\ti 10^8$         &   22.92                               & & 8.0  &                   \\ 
1222+22            &  $8.3\ti 10^7$           & $0.9 \ti 0.4$          & $3.5\ti 10^8$         &   24.99                               & & 19.4 &                   \\ 
1309+355           &  $2.5\ti 10^8$           & \mc{1}{c}{--}          &                       &   23.30                               & & 4.9  &                   \\ 
1351+640           &  $2.4\ti 10^7$           & \mc{1}{c}{--}          &                       &   21.60                               & & 2.4  &                   \\ 
1407+26            &  $1.1\ti 10^9$           & $0.7 \ti 0.4$          & $8.6\ti 10^9$         &   25.56                               & & 24.3 &                   \\ 
1700+518           &  $1.0\ti 10^7$           & \mc{1}{c}{--}          &                       &   22.45                               & & 11.3 &                   \\ 
2209+184           &  $9.7\ti 10^8$           & \mc{1}{c}{--}          &                       &   23.09                               & & 2.8  &                   \\ 
\hline\hline
\end{tabular}
\parbox{162mm}{\caption[sourcetable]{\label{tab:der_results} Derived
physical parameters from the VLBA observations.  Column 2 shows the
brightness temperatures derived according to the formula quoted in the
main text, using the sizes of the synthesised beam and total VLBA flux
densities listed in Table 2.  Where our snapshot data permitted robust
and well-constained deconvolved sizes to the emitting regions to be
fitted, we tabulate those in column 3, together with re-derived
brightness temperatures using these deconvolved sizes and the peak
VLBA flux densities. Also listed are the luminosities (in ${\rm W\,
Hz^{-1} sr^{-1}}$) at 8.4 GHz of the compact emission of the RQQs we
detected together with the characteristic size of the emitting region
(this is the geometric mean of the major and minor axes of the
synthesised beam converted into a physical distance in parsecs within
the assumed cosmology). }}
\end{table*}
\normalsize

Table~\ref{tab:der_results} shows that the RQQs require central
engines which can supply luminosities of $10^{21.6}$ -- $10^{25.6}$
${\rm W\, Hz^{-1} sr^{-1}}$ arising from regions of a few cubic parsec.
The peak luminosity at 5 GHz of the most powerful supernova known
(1986J) (Rupen et al.\ 1987) is $\sim 10^{20} {\rm W\, Hz^{-1}
sr^{-1}}$ so between 10 and 1000 ($10^5$ for the two high redshift,
very luminous objects) of these close to peak luminosity would thus be
required to power a single RQQ.  This would require a very sustained
supernova rate with an unprecendented supernova space density: the
(conservatively derived) values for the size of the emitting region
yield volumes between $10^5$ -- $10^7$ times smaller than in the
starburst model of Terlevich and Boyle (1993) or observed in the M82
starburst galaxy (Muxlow et al.\ 1994).  The brightness temperature
quantifies this by consideration of the flux emanating from a given
solid angle --- thus radio emission powered by a starburst would not
be expected to have a high brightness temperature because of the
spatial separation expected for the supernovae.

These results strongly suggest that for these radio-quiet quasars
which we detected, their radio emission is {\em not} dominated by
starbursts and imply that they have central engines similar to those
in RLQs, but producing only weak radio jets.

In a recent study (Falcke, Patnaik \& Sherwood 1996) high brightness
temperatures were found for three RIQs (which are common to our
sample) based on fits to the emission region size deconvolved from the
synthesised beam of the telescope. Our observations being considerably
shorter do not in all cases allow us to similarly obtain robust, well
constrained, fits to the sizes of the emission-regions.

\subsection{When should a quasar be deemed radio-quiet?}

We return to the question of when a quasar should be appropriately
classified as radio-loud. Of the three classifications outlined in the
Section 2, none considers the contribution to the total radio emission
from cores which might be Doppler boosted, if the cores are indeed
powered by relativistic jets. The bimodality in radio-loudness would
undoubtedly be more pronounced if instead of comparisons based on
total radio luminosity, the `total minus core' radio luminosity,
(i.e., only the contribution from unbeamed emission), were to be used
(see e.g., Kukula et al.\ 1998). It is imperative to ascertain whether
the radio flux densities quoted from the literature for those quasars
believed to be non-radio-loud, which we have detected with the VLBA,
are representing the extent of the radio-emission from these objects,
and thus that our detections are of genuinely radio-quiet quasars. We
thus checked the NVSS survey (Condon 1994) (VLA D-array 1.4-GHz maps),
the 6C (Hales et al.\ 1988, Hales et al.\ 1990) and the 7C (Waldram et
al.\ 1996) surveys at 151 MHz for evidence of diffuse extended lobe
emission related to these RQQs\footnote{1222+22 and 1407+26 are in
areas of sky covered by the 7C survey (Waldram et al.\ 1996).
1309+355 is in an area of sky covered by the 6C-II survey (Hales et
al.\ 1988).  1700+518 and 1351+640 are in an area of sky covered by
6C-III (Hales et al.\ 1990).}.  For the 6C and 7C surveys the {\sc
rms} background measurements are roughly $\sim$ 25 -- 50 mJy/beam in
the absence of any confusion. For the NVSS survey the {\sc rms}
background is $\sim 0.2$ mJy/beam.  We found {\em no} evidence of any
related extended lobe emission for our detected RQQs. We therefore
believe that all of the objects we detected originally classified as
radio-quiet are correctly classifed as such, although their
radio-quietness would be more dramatically evident were the criteria
to be based on extended radio emission only.

\subsection{What are the boosted counterparts of RQQs?}
\label{sec:riq}
One important question is whether the jets in RQQs are indeed
relativistic near the central engine (as in RLQs), albeit with a much
lower bulk kinetic power. If so a subset of the RQQ population, namely
those whose jet axes are oriented close to our line-of-sight, would be
expected to exhibit Doppler boosted emission. Such a scenario was
first proposed by Miller et al.\ (1993) based on a study of the [OIII]
luminosity versus radio luminosity plane for a sample of optically
selected quasars (those quasars from BQS with $z < 0.5$). They found
that radio-loud objects exist only at high [OIII] luminosity and that
for RQQs there is a tight correlation between the radio and the [OIII]
luminosity. A number of objects in the radio-quiet region of the plot
did not lie so tightly on this correlation; Miller et al.\ suggested
that their location on the plane could be explained if their radio
emission was Doppler boosted, i.e., they were the beamed counterparts
of RQQs (with Lorentz $\gamma \sim 5$); they termed such objects
radio-intermediate quasars. If the criteria for radio-loudness is
based on extended emission alone (as discussed in Section 4.2) then
the RIQs are clearly members of the RQQ population.  The cores of RIQs
might be expected to have higher $T_{\rm B}$ than objects whose radio
emission is not Doppler boosted.  While there is no such clear
correlation from the numbers in this small sample, we note that the
comparison of brightness temperatures is an important tool in testing
Miller et al.'s hypothesis.

A number of the sources we detected are among those deemed by Miller
et al.\ as RIQs. It is therefore conceivable that the RIQs represent
those RQQs with a jet-producing central engine while the true
radio-quiet objects lack such a central engine --- but the dichotomy
posed at the beginning of the paper is little changed, as it is still
necessary to explain why the RIQ population cannot form the powerful
radio jets seen in the RLQ population, even though there is now direct
evidence that some contain jet-producing central engines.

\subsection{Conclusions}

Our results have shown that some radio-quiet quasars show evidence for
a central engine resembling those in radio-loud quasars; the evidence
we present is consistent with the sample objects being boosted
examples of a homogeneous population of radio-quiet quasars with
relativistic jets.  Our study underlines the need to address the
important question of why powerful radio jets are not seen in RQQs
even though a significant fraction of their central engines possess
the essential characteristics of those in RLQs; we note that there
have been various suggested explanations of this, including for
example, lack of black-hole spin (Blandford 1993) or the necessary
presence of a hot atmosphere around the nucleus (Fabian \& Rees 1995).

\section*{Acknowledgments}

The VLBA and VLA are facilities of the National Radio Astronomy
Observatory, which is operated by Associated Universities, Inc., under
co-operative agreement with the US National Science Foundation.  This
research has made use of the NASA/IPAC Extragalactic Database (NED)
which is operated by the Jet Propulsion Laboratory, California
Institute of Technology, under contract with the National Aeronautics
and Space Administration. We thank Dr Craig Walker \& Dr Steve
Rawlings for helpful discussions, Dr Stephen Blundell for a careful
reading of the manuscript and the referee, Dr Robert Laing, for
constructive comments.

\end{document}